\documentclass[journal,comsoc]{IEEEtran}

\usepackage{cite}

\usepackage{amsmath,amssymb,amsfonts}
\usepackage{algorithmic}
\usepackage{graphicx}
\usepackage{textcomp}
\usepackage{xcolor}
\usepackage{float}
\usepackage[nolist]{acronym}
\usepackage[none]{hyphenat}
\let\labelindent\relax
\usepackage[shortlabels]{enumitem}
\usepackage{tabularx}
\usepackage{comment}
\usepackage[normalem]{ulem}
\usepackage{url}

\usepackage{cancel}

\usepackage{placeins}

\def\BibTeX{{\rm B\kern-.05em{\sc i\kern-.025em b}\kern-.08em
    T\kern-.1667em\lower.7ex\hbox{E}\kern-.125emX}}

\newif\ifrev
\revfalse % for false
\ifrev

\newcommand{\luirev}[1]{\textcolor{blue}{#1}}

\newcommand{\alvrev}[1]{\textcolor{blue}{#1}}

\else
    
\newcommand{\luirev}[1]{#1}

\newcommand{\alvrev}[1]{#1}

\fi
\begin{document}

\title{Spreading Factor Assignment Strategy for Coverage and Capacity Flexible Tradeoff}

\author{Luiz~Q.~R.~da~C.~Filho,~Alvaro~A.~M.~de~Medeiros,~J\'{e}ssika~C.~da~Silva,~Vicente~A.~de~Sousa~Jr.~and~N\'{i}bia~S.~Bezerra
\thanks{Luiz Filho and Alvaro Medeiros are with Federal University of Juiz de Fora (e-mails: \{luiz.quirino, alvaro\}@engenharia.ufjf.br.). Jéssika Silva e Vicente~Sousa are with Federal University of Rio Grande do Norte, Brazil (e-mails: \{jessie,vicente.sousa\}@ufrn.edu.br). N\'{i}bia Bezerra is with Lule{\aa} University of Technology,
Sweden (e-mails: nibia.souza.bezerra@ltu.se). This study was financed in part by the Coordena\c{c}\~{a}o de Aperfei\c{c}oamento de Pessoal de N\'{i}vel Superior - Brasil (CAPES) - Finance Code 001. The proof of concept simulations provided by this Letter was supported by High Performance Computing Center~(NPAD/UFRN).} \thanks{Digital Object Identifier: 10.14209/jcis.2022.5}
}

% \hrule{\columnwidth}

\maketitle

%\markboth{XXXVII SIMPÓSIO BRASILEIRO DE TELECOMUNICAÇÕES E PROCESSAMENTO DE SINAIS - SBrT2019, 29/09/2019--02/10/2019, PETRÓPOLIS, RJ}{XXXVII SIMPÓSIO BRASILEIRO DE TELECOMUNICAÇÕES E PROCESSAMENTO DE SINAIS - SBrT2019, 29/09/2019--02/10/2019, PETRÓPOLIS, RJ} 

\begin{acronym}
  \acro{1G}{First Generation}
  \acro{2G}{Second Generation}
  \acro{3G}{Third Generation}
  \acro{4G}{Fourth Generation}
  \acro{5G}{Fifth Generation}
  \acro{5GRA}{Remote Areas applications}
  \acro{5GPHY}{5G Physical Layer}
  \acro{ADC}{analogic-to-digital converter}
  \acro{AGC}{automatic gain control}
  \acro{ASIP}{Application Specific Integrated Processors}
  \acro{AWGN}{additive white Gaussian noise}
  \acro{BDTM}{burst data transfer mode}
  \acro{BER}{bit error rate}
  \acro{BS}{base station}
  \acro{CDTM}{continuous data transfer mode}
  \acro{CFO}{Carrier Frequency Offset}
  \acro{CHF}{Characteristic Function}  
  \acro{CoMP} {Cooperative Multi-point}
  \acro{CP}{cyclic prefix}
  \acro{CR}{Cognitive Radio}
  \acro{CS}{cyclic suffix}
  \acro{CSI}{channel state information}
  \acro{CSMA}{carrier sense multiple access}
  \acro{DFT}{discrete Fourier transform}
  \acro{DPD}{digital pre-distortion}
  \acro{DZT}{discrete Zak transform}
  \acro{eMBB}{enhanced mobile broadband}
  \acro{EPC}{evolved packet core}
  \acro{FBMC}{Filter-bank multi-carrier}
  \acro{FDE}{frequency-domain equalizer}
  \acro{FDMA}{frequency division multiple access}
  \acro{FD-OQAM-GFDM}{frequency-domain OQAM-GFDM}
  \acro{FEC}{forward error control}
  \acro{FPGA}{Field Programmable Gate Array}
  \acro{FTN}{Faster than Nyquist}
  \acro{FT}{Fourier transform}
  \acro{FSC}{frequency-selective channel}
  \acro{GFDM}{Generalized Frequency Division Multiplexing}
  \acro{GS-GFDM}{guard-symbol GFDM}
  \acro{HPA}{high power amplifier}
  \acro{IBI}{inter-block interference}  
  \acro{ICI}{inter-carrier interference}
  \acro{IDFT}{Inverse Discrete Fourier Transform}
  \acro{IFI}{inter-frame interference}
  \acro{IMS}{IP multimedia subsystem}
  \acro{IoT}{Internet of Things}
  \acro{IP}{Internet Protocol}
  \acro{IQ}{in-phase and quadrature}
  \acro{ISI}{inter-symbol interference}
  \acro{IUI}{inter-user interference}
  \acro{KPI}{key performance indicator}
  \acro{LDPC}{low density check parity code}
  \acro{LLR}{log-likelihood ratio}
  \acro{LMMSE}{linear minimum mean square error}
  \acro{LTE}{Long-Term Evolution}
  \acro{LTE-A}{Long-Term Evolution - Advanced}
  \acro{M2M}{Machine-to-Machine}
  \acro{MA}{multiple access}
  \acro{MAC}{medium access control layer}
  \acro{MF}{Matched filter}
  \acro{MIMO}{multiple-input multiple-output}
  \acro{MMSE}{minimum mean square error}
  \acro{MRC}{maximum ratio combiner}
  \acro{MSE}{mean-squared error}
  \acro{mMTC}{massive machine type communication}
  \acro{MTC}{machine type communication}
  \acro{MU}{multi user}
  \acro{NEF}{noise enhancement factor}
  \acro{NFV}{network functions virtualization}
  \acro{OFDM}{Orthogonal Frequency Division Multiplexing}
  \acro{OOB}{out-of-band}
  \acro{OQAM}{Offset Quadrature Amplitude Modulation}
  \acro{PAPR}{peak to average power ratio}
  \acro{PHY}{physical layer}
  \acro{PRBS}{Pseudo Random Bit Sequence}
  \acro{PSD}{Power Spectrum Density}
  \acro{QAM}{quadrature amplitude modulation}
  \acro{QPSK}{quadrature phase shift keying}
  \acro{QoE}{Quality of Experience}
  \acro{QoS}{Quality of Service}
  \acro{RC}{raised-cosine}
  \acro{RF}{radio frequency}
  \acro{ROF}{roll-off factor}
  \acro{RRC}{root raised cosine}
  \acro{SC}{single carrier}
  \acro{SC-FDE}{Single Carrier Frequency Domain Equalization}
  \acro{SC-FDMA}{Single Carrier Frequency Domain Multiple Access}
  \acro{SCD}{Successive Cancellation Decoding}
  \acro{SDN}{software-defined network}
  \acro{SDR}{software-defined radio}
  \acro{SDW}{software-defined waveform}
  \acro{SEP}{symbol error probability}
  \acro{SER}{symbol error rate}
  \acro{SIC}{successive interference cancellation}
  \acro{SISO}{single-input single-output}
  \acro{SMS}{Short Message Service}
  \acro{SNR}{signal-to-noise ratio}
  \acro{ST}{space-time}
  \acro{STO}{Symbol Timing Offset}
  \acro{STC}{space time code}
  \acro{STFT}{short-time Fourier transform}
  \acro{TD-OQAM-GFDM}{time-domain OQAM-GFDM}
  \acro{TR-STC}{time-reversal space-time coding}
  \acro{TR-STC-GFDMA}{TR-STC Generalized Frequency Division Multiple Access}
  \acro{TVC}{time-variant channel}
  \acro{TVWS}{TV white space}
  \acro{UHF}{Ultra High Frequency}
  \acro{URLL}{ultra-reliable low latency}
  \acro{V2V}{vehicle-to-vehicle}
  \acro{VHF}{Very High Frequency}
  \acro{V-OFDM}{Vector OFDM}
  \acro{ZF}{zero-forcing}
  \acro{W-GFDM}{windowed GFDM}
  \acro{WHT}{Walsh-Hadamard Transform}
  \acro{WLAN}{wireless Local Area Network}
  \acro{WLE}{widely linear equalizer}
  \acro{WLP}{wide linear processing}
  \acro{WRAN}{Wireless Regional Area Network}
  \acro{WSN}{wireless sensor networks}
  \acro{TLS}{transport layer security}
  \acro{BRICS}{Brazil, Russia, India, and South Africa}
  \acro{LPWAN}{low-power wide-area network}
  \acro{UE}{user equipment}
  \acro{GPS}{Global Positioning System}
  \acro{RFID}{Radio-frequency identification}
  \acro{CAN1}{Controller Area Network 1}
  \acro{CAN2}{Controller Area Network 2}
  \acro{P2P}{Peer-to-peer}
  \acro{ICIC}{inter-cell interference coordination}
  \acro{ACK}{acknowledgment}
\end{acronym}

\begin{abstract}
%The Internet of Things concept has been looking for a solution for the connection of dense sensor networks, LoRa is a physical layer technology that is gaining popularity due to its ability to connect multiple devices in a wide area of coverage, with a low power consumption and with sufficient resistance to interference, these characteristics are due to its modulation based on chirp spread-spectrum in which it produces a rather robust signal and with different configurations that are orthogonal to each other. In this article we investigate the efficiency of LoRa to send multiple uplink traffic, to evaluate the efficiency we use different ways of allocating the spread factor in different scenarios, the results confirm that using different spread factors can increase the number of users transmitting.
%A dense wireless network for IoT applications needs an efficient transmission solution. 
LoRa is a physical layer technology with the ability to connect multiple devices in a wide area of coverage, with low power consumption and with interference robustness. \alvrev{The LoRaWAN specification introduces the protocol for communication between multiple devices and the gateway and defines an algorithm for spreading factor allocation. } %Its modulation, based on chirp spread-spectrum, produces a rather robust signal based on orthogonal spread factors.
In this Letter, we investigate the efficiency of LoRa to send multiple uplink streams, \alvrev{analyzing different spreading factor allocation strategies to bring light to the coverage-capacity tradeoff. Additionally, we present a complete open-source simulation framework based on ns-3 simulator that can be used to propose, test and analyze the performance of new algorithms or heuristics that may outperform LoRaWAN ADR or any other baseline strategies defined here.}  %The results confirm the dependency between the system capacity and the spread factor allocation strategy. 

%These characteristics are due to its modulation based on chirp spread-spectrum, which produces a rather robust signal based on orthogonal spread factors. In this Letter, we investigate the efficiency of LoRa to send multiple uplink traffics, considering different ways of allocating the spreading factor in different scenarios. The results confirm the dependency of the system capacity and the spread factor allocation strategy. 
\end{abstract}


\section{Introduction}

\alvrev{A well-known  Low-Power Wide Area Network (LPWAN) technology is the LoRaWAN~\cite{lpwansurvey}, which is based on the physical layer technique LoRa (\emph{Long Range}) in order to  guarantee a robust signal over  noise and interference, even in wider area deployments.} LoRa technology, which is based on Chirp Spread-Spectrum (CSS), uses signals with different Spreading Factors ($SF$s) orthogonal to each other, so that it is possible to choose different factors for different devices at the same location, reducing interference between devices transmitting simultaneously. A higher $SF$ is allocated to End Devices (ED) located far from the Gateway (GW) enhancing robustness instead of throughput. A lower $SF$  can be allocated to EDs closer to GW, in order to increase throughput and reduce the frequency bandwidth occupancy (air time). \alvrev{Therefore, the $SF$ allocation plays a major role in the LoRaWAN network operation. }

The LoRaWAN specification determines a policy for $SF$ assignment for the devices, namely Adaptive Data Rate (ADR), based on the ED's received power level: the higher the received power level is, the lower the $SF$ value is assigned.

Due to the growing interest in LoRaWAN, several studies~\cite{an1,an2,an3,an4} made an analytical performance evaluation, and the LoRaWAN perfomance is also analyzed via real network tests \cite{real1,real2,real3,real4} and using ns-3 \cite{surveyNs-3Jcis,adrns3,perfomanceeva,difconf,jakarta}, OMNeT++ \cite{explora,omnet2} and LoRaWANSim \cite{adrlorawansim} simulators.

In this Letter, we present a flexible strategy for $SF$ assignment that can be configured for different network scenarios. The simplicity of such strategy allows a direct implementation on equipment with hardware constraints and an interesting analysis considering both viewpoints: capacity and coverage.  We also compare our proposal with the ADR algorithm and other allocation policies that are focused only on either capacity or coverage. %For this, we perform our evaluation using the ns-3 simulator. 
Our key contributions are \alvrev{the following:}%: (i) a simple and generic $SF$ allocation strategy that can be configured to increase capacity or coverage, and (ii) an open and available framework for both capacity and coverage performance evaluation based on ns-3~\cite{github}. %\footnote{Publicly available at \url{https://github.com/vicentesousa/ns-3-lora-gppcom}.}.
%\begin{itemize}
    %\item  a simple and generic SF assignment strategy that can be configured to increase capacity or coverage performance of ADR algorithm, and
 %   \item  a simple and generic ADR strategy that can be configured to increase capacity or coverage, and
 %   \item an open and available framework for both capacity and coverage performance evaluation based on ns-3\footnote{Publicly available at \url{https://github.com/vicentesousa/ns-3-lora-gppcom}.}.
%\end{itemize}
\alvrev{\begin{itemize} 
 \item a simple and generic $SF$ allocation strategy that can be configured to increase capacity or coverage, and 
 \item an open and available framework for both capacity and coverage performance evaluation based on ns-3~\cite{github}.
 \end{itemize}}
 
\alvrev{This Letter is organized as follows. %Section~\ref{LoRa} describes how the LoRa and LoRaWAN ADR works. 
Section~\ref{SF} briefly introduces LoRa and LoRaWAN, and presents the $SF$ allocation methods evaluated in this Letter. Section~\ref{simulacao} describes the simulation scenarios, and Section~\ref{resultados} analyzes the obtained results. The final considerations are presented in Section~\ref{conclusao}.}

\section{Spreading Factor Assignment Methods}
\label{SF}

LoRa is a technology patented by Semtech Corporation~\cite{semtech}. The term LoRaWAN refers to the architecture and specification of MAC layer, developed by the LoRa Alliance. Although the LoRa technology is proprietary, the LoRaWAN protocol is open source~\cite{lorasurvey}, and specifies parameters such as frame format, device classes and security protocols~\cite{lorawan_std}.

The LoRa modulation, which is based on a variation of CSS modulation, as so as the operation in sub 1 GHz bands, are responsible for the high range and resistance to interference and noise. Its bit rate~$R_b$ is defined as %$R_b = {SF}/{T_{s}} = SF.{BW}/{2^{SF}}$,
\begin{equation}
\label{eq4}
R_b = \frac{SF}{T_{s}} = SF \frac{BW}{2^{SF}},
\end{equation}
where $SF$ is the Spreading Factor, which determines the number of bits in the symbol and can vary from~7 to~12. In addition, $SF$ determines the duration of a symbol~$T_s$, which is the time required to scan the entire band of the signal~$BW$. Although a higher $SF$ signal has a lower data rate, it is more tolerant to interference or noise due to easier detection at the receiver. Thus, the choice of~$SF$ represents a tradeoff between transmission range and data rate.

%\begin{comment}

\alvrev{LoRaWAN standards also defines the Adaptive Data Rate (ADR) algorithm that manages the power and data rate of the ED, by assigning its $SF$~\cite{lorawan_std}. After a configurable number of packages, the ED sends a request for an acknowledgement (ACK). If it does not receive the ACK, it tries to reestablish the connection by increasing the power and the $SF$, while reducing the data rate.}

Thus, as each Spreading Factor requires a minimum received power to operate properly, the ADR algorithm consider the receiver sensitivity for each $SF$ before its assignment. Table~\ref{table_sf} presents the receiver sensitivity  and the bit rate for each $SF$~\cite{semtech}. The larger the quantity of EDs operating with the same $SF$, the higher is the probability of interference, and consequently, the packet error \alvrev{rate.} %However, due to the fact that two different $SF$s do not interfere with each other, EDs may also be allocated to a higher or a lower $SF$ independently of its receiver sensitivity, in order to increase the efficiency of the network.  

\begin{table}[htb]
%\centering
\caption{Sensitivity and Bit Rate for each $SF$.}
\begin{center}
\label{table_sf}
\begin{tabular}{ |c|c|c|}
\hline
\textbf{SF}& \textbf{Sensitivity (dBm)} & \textbf{Bit rate (kb/s)} \\ \hline 

\textbf{7} &  -123 &  5.468 \\
\hline 

\textbf{8} & -126 & 3.125 \\
\hline 

\textbf{9} & -129 & 1.757 \\
\hline 

\textbf{10} & -132 & 0.976 \\
\hline 

\textbf{11} &  -134.5 & 0.537 \\
\hline 

\textbf{12} &  -137 &  0.293\\
\hline 

\end{tabular}
\end{center}
\end{table}

%\end{comment}

In this Letter, we analyze simple $SF$ assignment algorithms that can be applied to EDs with modest processing capabilities and may not generate extra traffic on the network. For this, we define the assignment vector $\mathbf{a}=\{a_7, a_8, ..., a_{12}\}$, where~$a_7$ denotes the fraction of total EDs that are allocated to SF~$7$, $a_8$ to SF~$8$, and so forth. Thus, we have that $\sum_{i=7}^{12}a_i=1$. Nine different methods of Spreading Factor allocation are evaluated:
\begin{itemize}%[label=(\Roman*)]
\item%[\textbf{(I)}]%\label{modo1} 
\textbf{Fixed at the lowest $SF$ (I)} -
All EDs use SF~$7$ (highest data rate), i.e. $a_7=1$ and $a_i=0~\forall~i\neq7$. %In this case, we have that $a_7=1$ and $a_i=0~\forall~i\neq7$. This is the case where the highest data rate is assigned for all users.

\item%[\textbf{(I)}] %\label{modo1.1} 
  \textbf{Fixed at the highest $SF$ (II)} -
All EDs use SF~$12$ (lowest data rate), i.e. $a_{12}=1$ and $a_i=0~\forall~i\neq12$.%In this case, we have that $a_{12}=1$ and $a_i=0~\forall~i\neq12$. Thus, all users have the maximum coverage guaranteed. 

\item%[\textbf{(III)}] %\label{modo2} 
\textbf{Equally divided (III)} -
The $N$ EDs are ranked by received power and equally divide into six groups in a way that the the first group is allocated to SF~$7$, the second to SF~$8$ and so forth ($a_i=\lfloor N/6 \rfloor$).

\item 
%\label{modo3} 
\textbf{Arbitrarily divided} -
The $N$ EDs are ranked and divided like method~(III), but the quantity of EDs per group is an arbitrary fraction given by the assignment vector~$\mathbf{a}$. We define two assignment vectors:%This strategy is evaluated considering two assignment vectors:

\begin{itemize}
    \item \textbf{Capacity enhancement (IV)} - The algorithm follows %allocates ED to each SF group so that 
    $\mathbf{a}=\{0.6, 0.2, 0.05, 0.05, 0.05, 0.05\}$.
    \item \textbf{Coverage enhancement (V)} - The algorithm follows %tries to allocate ED to each SF group so that 
    $\mathbf{a}~=~\{0.05, 0.05, 0.05, 0.05, 0.2, 0.6\}$.
\end{itemize}

%
%The devices are divided just like the previous method, placing the devices with the best signal with the lowest ~$SF$s, but due to the fact that the lower the Spreading Factor the higher the data rate, more devices are placed in the group with minors Spreading Factors. Thus, SFs are allocated so that 60\% of devices, with the best signal, are with Factor equal to~7, 20\% equal to~ 8, 5\% equal to~9, 5\% equal to~10, 5\% equal to~11 and the~5\% with the worst signal have SF equal to~12.
%
\item 
%\label{modo4} 
\textbf{Sensitivity based (VI)} -
%In this allocation mode, 
Each ED has the lowest $SF$ possible, so that the power received from that ED is greater than its sensitivity. This policy corresponds to ADR defined in LoRaWAN standards and the assignment vector depends on the deployment factors, such as position of EDs, wireless channel, and interference.
\item 
%\label{modo5} 
\textbf{Sensitivity based arbitrarily divided} - This method follows the Arbitrarily divided one, %The EDs are allocated in each SF group in the same way as in Arbitrarily divided method, 
but respecting the sensitivity rule defined in method~(VI). This strategy affords $SF$ allocations that may differ from LoRaWAN ADR algorithm, since some EDs may be assigned to higher or lower SFs, improving coverage or capacity, respectively. %This strategy provides an interesting investigation: as allocation of EDs on higher SFs increases, so does the network coverage and it may also reduces interference on lower SFs, which, on the other hand, may also increases capacity. 
We evaluate this strategy using two assignment vectors: 

\begin{itemize}
    \item \textbf{Capacity enhancement (VII)} - The algorithm follows %tries to allocate ED to each SF group so that 
    $\mathbf{a}=\{0.6, 0.2, 0.05, 0.05, 0.05, 0.05\}$, respecting the sensitivity rule defined by ADR.
    \item \textbf{Coverage enhancement (VIII)} - The algorithm follows  %tries to allocate ED to each SF group so that 
    $\mathbf{a}~=~\{0.05, 0.05, 0.05, 0.05, 0.2, 0.6\}$, respecting the sensitivity rule defined by ADR.
\end{itemize}

\item
\textbf{Randomly assigned (IX)} - Each ED chooses a $SF$ randomly from all six SFs with the same probability. %Due to its simplicity and randomness, this strategy is a performance lower bound of our comparison.

%\item
%\textbf{Randomly assigned (X)} - 
\end{itemize}

\section{Simulation Scenarios}
\label{simulacao}

%In order to evaluate the SF allocation methods, we use version 3.29 of the ns-3~\cite{ns3}. %. The ns-3 simulator is an open source discrete event simulator written in C++ that is now available in version 3.29~\cite{ns3}. 
We use the version 3.29 of the ns-3~\cite{ns3} for our evaluation, and the module developed by D. Magrin in~\cite{lora1}, with a single GW multi-EDs deployment. \alvrev{There is no native LoRaWAN module for ns-3 and different modules can be used. Please refer to~\cite{surveyNs-3Jcis} for a more detailed description about the module we used and others proposed for ns-3.}

%There are several implementations of LoRa modules for ns-3 from different research groups available. In this paper, we use the module   developed by D. Magrin in~\cite{lora1}. % as part of his master's thesis by the University of Padova in Italy. 
%The simulations are configured to evaluate traffic from multiple EDs to a single GW.

In this way, a GW is placed surrounded by several EDs that send a packet in a random time within an interval of~10 minutes. Table~\ref{table:cenarioreal} presents the parameters of simulations based on smart city scenario~\cite{perfomanceeva}. \alvrev{The log-distance path loss propagation model~\cite{rappaport2002wireless} was adopted so that we could focus on the capacity-coverage trade-off without randomness imposed by more complex channel models. However, since such channel models are defined in ns-3, this work can be extended to analyze the channel effects on the $SF$ allocation.}
%A survey on  LoRaWAN implementations on ns-3 can be found in~\cite{surveyNs-3Jcis}.

\alvrev{The LoRaWAN module on ns-3 adopted in this work  is not fully implemented, and some MAC layer signaling is missing. As the $SF$ update is ideal, the results presented are considered for a steady-state network. There is no intra-$SF$ orthogonality. Collisions inter-$SF$ are handled in a quasi-orthogonal way, i.e. they can occur for high received power difference between simultaneous signals. Since the  \emph{LinkADRReq} command~\cite{lorawan_std} is not implemented of this module, the configuration of transmission power and $SF$ is considered to
be ideal.}

%\alvrev{Teremos sombreamento? Se sim, texto aqui. O que e feito com os pacotes que restam no final da simulação? São contados como pacotes perdidos?}

%\vspace{-.3cm}
\begin{table}[htb]
\caption{Simulation Parameters.}
\begin{center}
    \begin{tabular}{ | c | c |}
    \hline
    \textbf{Parameter}& \textbf{Value}\\ \hline   
    %Number of GWs & 1  \\ \hline
    
    Number of EDs & 1000, 2000, 3000, 4000, 5000, 6000 \ref{cenario1} \\
                     & 1000, 3000, 6000 \ref{cenario2} \\  \hline
    
    Packet size &  23 Bytes \\ \hline
    
    Radius of circle &  3000, 6000, 10000~m  \ref{cenario1}\\
                     &  2000, 3000, 4000, 6000, 8000, 10000~m \ref{cenario2} \\ \hline
    
    %Number of packets &  1 \\ \hline
    
    Traffic direction & only uplink  \\ \hline
    
    Bandwidth & 125~KHz \\ \hline
    
    %Propagation model & Log-distance \\ \hline
    
    Path loss values  & $d_0$=1~m; $\overline{PL}(d_0)$=7.7; $n$=3.7 \\ \hline
    
    Transmission power & 14~dBm\\ \hline
    
     \luirev{Number of simulation runs} & \luirev{10}\\ \hline

    %Retransmissions & disabled \\ \hline
    
   %Number of simulations &  20\\ \hline

    \end{tabular}
\end{center}
\label{table:cenarioreal}
\end{table}

We define two sets of simulations, each one composed of three campaigns, to analyze the methods defined in Section~\ref{SF}:
\begin{enumerate}[label=(\Alph*)]
\item \label{cenario1} \textbf{Capacity analysis} -
Simulation campaigns are carried out
with EDs uniformly distributed in a circle of radius of 3, 6 and 10 km, and the number of EDs increases as presented in %from 1000 to 6000
Table~\ref{table:cenarioreal}.% in order to observe the performance of the strategies as the number of EDs increases.
\item \label{cenario2} \textbf{Coverage analysis} -
Simulation campaigns are carried out with the number of EDs fixed at 1000, 3000 and 6000, and the radius of the circular area where they are uniformly distributed increases as shown in Table~\ref{table:cenarioreal}. %for each simulation with the purpose of evaluating how each allocation method is influenced by the ED's positions and the GW's received power.
%(se os dispositivos estão muito juntos ou muito separados).OK
\end{enumerate}

%Our proposed SF assignment strategy \ref{modo5} is evaluated considering two assignment vector $\mathbf{a}$:
%\begin{itemize}
%    \item Capacity enhancement - The algorithm tries to allocate ED to each SF group so that $\mathbf{a}=\{0.6, 0.2, 0.05, 0.05, 0.05, 0.05\}$, respecting the sensitivity rule defined by ADR. This strategy is referred in the results as (VI-Cap).

%    \item Coverage enhancement - The algorithm tries to allocate ED to each SF group so that $\mathbf{a}~=~\{0.05, 0.05, 0.05, 0.05, 0.2, 0.6\}$, respecting the sensitivity rule defined by ADR. This strategy is referred in the results as (VI-Cov).
%\end{itemize}
%Method \ref{modo3} is also tested using  $\mathbf{a}=\{0.6, 0.2, 0.05, 0.05, 0.05, 0.05\}$. 

%Each simulation was performed twenty times, and the result presented is an average of the values obtained in each simulation. \textcolor{red}{esta informação pode ir para a tabela tambem}

\section{Results}
\label{resultados}
\begin{figure} [!hb]
\centering
\begin{minipage}[c]{0.9\columnwidth}
\includegraphics [width=\textwidth] {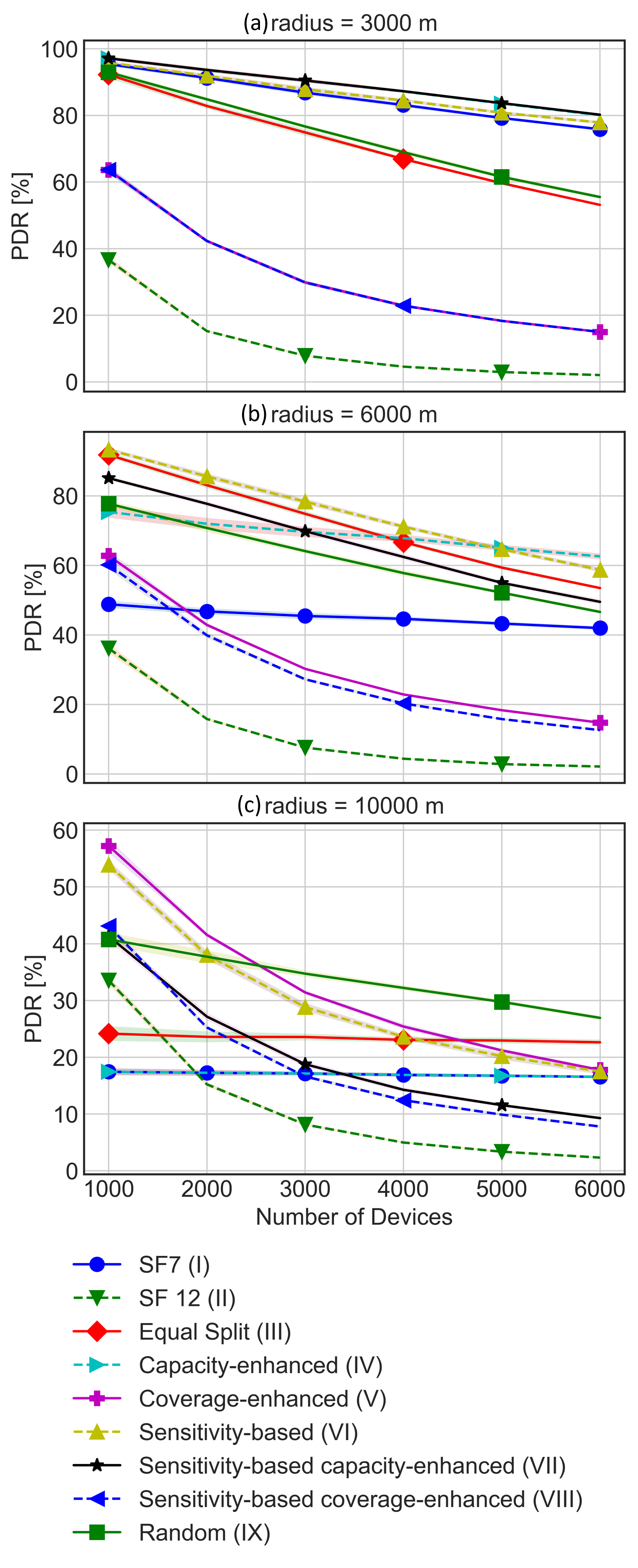}
\caption {\alvrev{Percentage of received packets in function of the number of EDs for different $SF$ allocation policies for EDs uniformly distributed in a circle of radius of 3~km (a), 6~km (b), and 10~km (c). }}
\label{fig:PacotesRecebidosCapacidade}
\end{minipage}
\end{figure}

\begin{figure} [!h]
\centering
\begin{minipage}[c]{0.9\columnwidth}
\includegraphics [width=\textwidth] {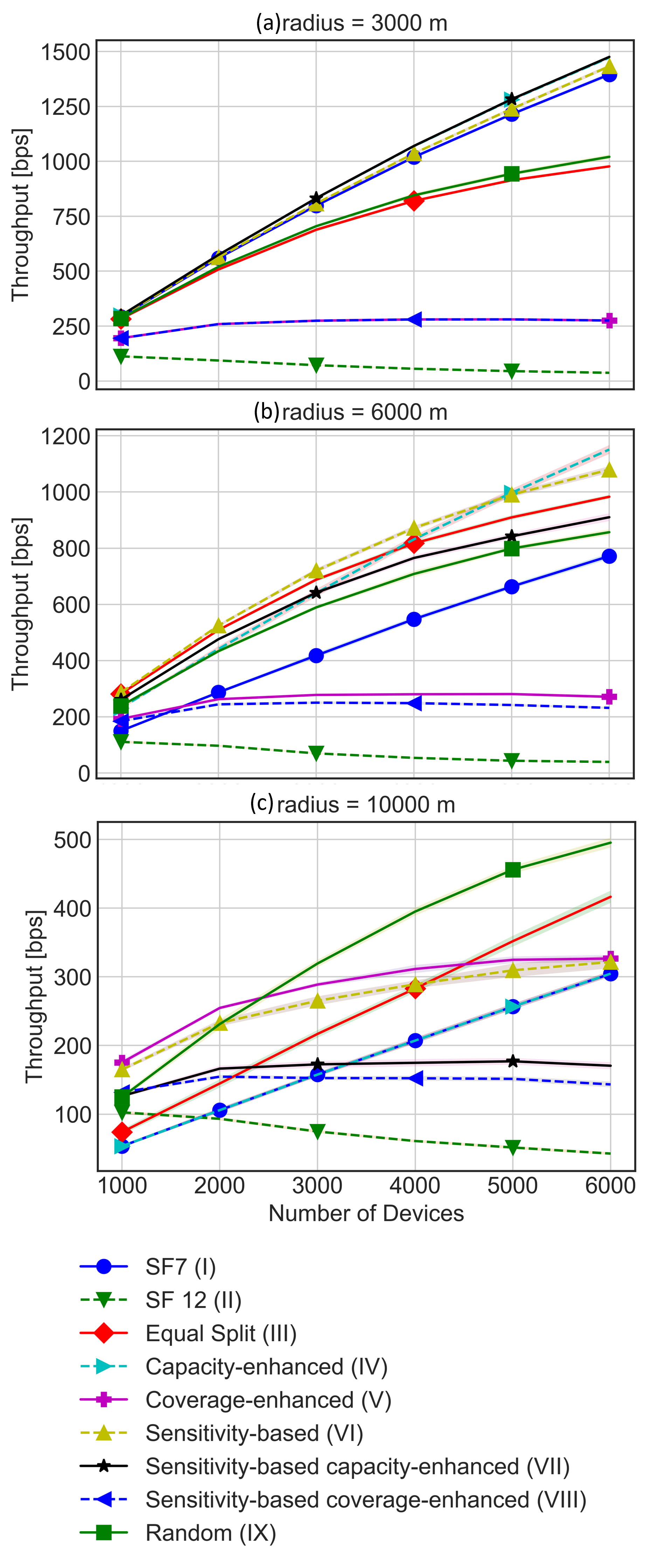}
\caption {\alvrev{Throughput of the network in function of the number of EDs for different $SF$ allocation policies for EDs uniformly distributed in a circle of radius of 3~km (a), 6~km (b), and 10~km (c). }}
\label{fig:VazaoCapacidade}
\end{minipage}
\end{figure}

%
%Figures~\ref{fig:PacotesRecebidosCapacidade} and~\ref{fig:VazaoCapacidade} present the results for the set of simulations~\ref{cenario1}, i.e. increasing the number of EDs. 
Figures~\ref{fig:PacotesRecebidosCapacidade} and~\ref{fig:VazaoCapacidade} show the percentage of correctly received packets and the throughput of all EDs, respectively, in function of the number of EDs for the $SF$ allocation methods presented in Section~\ref{SF}. 
\alvrev{The 95\% confidence interval is represented by the shadow around the curves.} We notice on Fig.~\ref{fig:PacotesRecebidosCapacidade}(a) and ~\ref{fig:VazaoCapacidade}(a) that for a smaller radius (3000~m), the coverage-based methods (II, V and~VIII) present the worst performance, since many EDs are allocated with higher $SF$s, increasing the probability of a collision (higher packet transmission time). Methods~(III) and~(IX) \alvrev{performs similarly}, as many EDs \alvrev{do not have} high~$SF$. Methods~(I) and~(VI) \alvrev{perform} better, since $SF$~7 provides the highest rates and the distance between EDs and the GW is \alvrev{low enough to prevent} packet loss. In the \alvrev{LoRaWAN ADR} method~(VI), most EDs are configured with~$SF$ 7, as the distance to GW is not high enough. The few EDs that operate below the sensitivity threshold are allocated to higher $SF$s, so that they can also transmit. Methods~(IV) and~(VII) \alvrev{outperform LoRaWAN ADR} and present the best performance since they allocate the EDs along all $SF$s with fewer collisions occurring in $SF$~7.
%##########
% Nova figura incluida por Vicente

%##########
% Nova figura incluida por Vicente

As the radius increases %to 6000m and 10000m
(Figures~\ref{fig:PacotesRecebidosCapacidade}(b), ~\ref{fig:PacotesRecebidosCapacidade}(c),~\ref{fig:VazaoCapacidade}(b), and~\ref{fig:VazaoCapacidade}(c)), we notice that the coverage-based algorithms improve their performance, since most of the EDs are allocated to high $SF$s, guarantying signal coverage and protection to interference. \alvrev{However, it can be observed from Figures~\ref{fig:PacotesRecebidosCapacidade}(c) and~\ref{fig:VazaoCapacidade}(c) that  strategies~(I), (III) and~(IV) increase the throughput and maintain their PDR level in the 10-km scenario. Although strategies~(I) and~(IV) focus on $SF$~7 to enhance capacity, strategy~(III) equally allocates EDs for all $SF$s. This indicates that inter-$SF$ interference might play a major role in the capacity-coverage trade-off.}

An interesting result is that methods~(III) and~(IX) outperform \alvrev{all methods including the LoRaWAN ADR method~(VI),} as both radius and number of EDs increase. This indicates that capacity and coverage can not be analyzed separately. As the number of EDs in a wider area increases, interference plays a major role. EDs with the same $SF$ tend to transmit with the same power, which means that a collision may be harmful for both packets. However, when EDs with different $SF$s transmits simultaneously, the difference of transmission power and air time may result in a successful reception for at least one of them. As method~(IX) outperforms method~(III), we verify that finding the optimum allocation vector is not as simple as dividing $SF$s equally. \alvrev{Furthermore, as the random allocation outperforms the LoRaWAN ADR method, improvements on the latter should consider other factors besides ED sensitivity.} In fact, our simulation framework available in~\cite{github} % \url{https://github.com/vicentesousa/ns-3-lora-gppcom} 
can leverage the investigation of algorithms to define vector $\mathbf{a}$, and consequently, allocation methods in LoRaWAN systems. 

\begin{figure} [!h]
\centering
\begin{minipage}[c]{0.9\columnwidth}
\includegraphics [width=\textwidth] {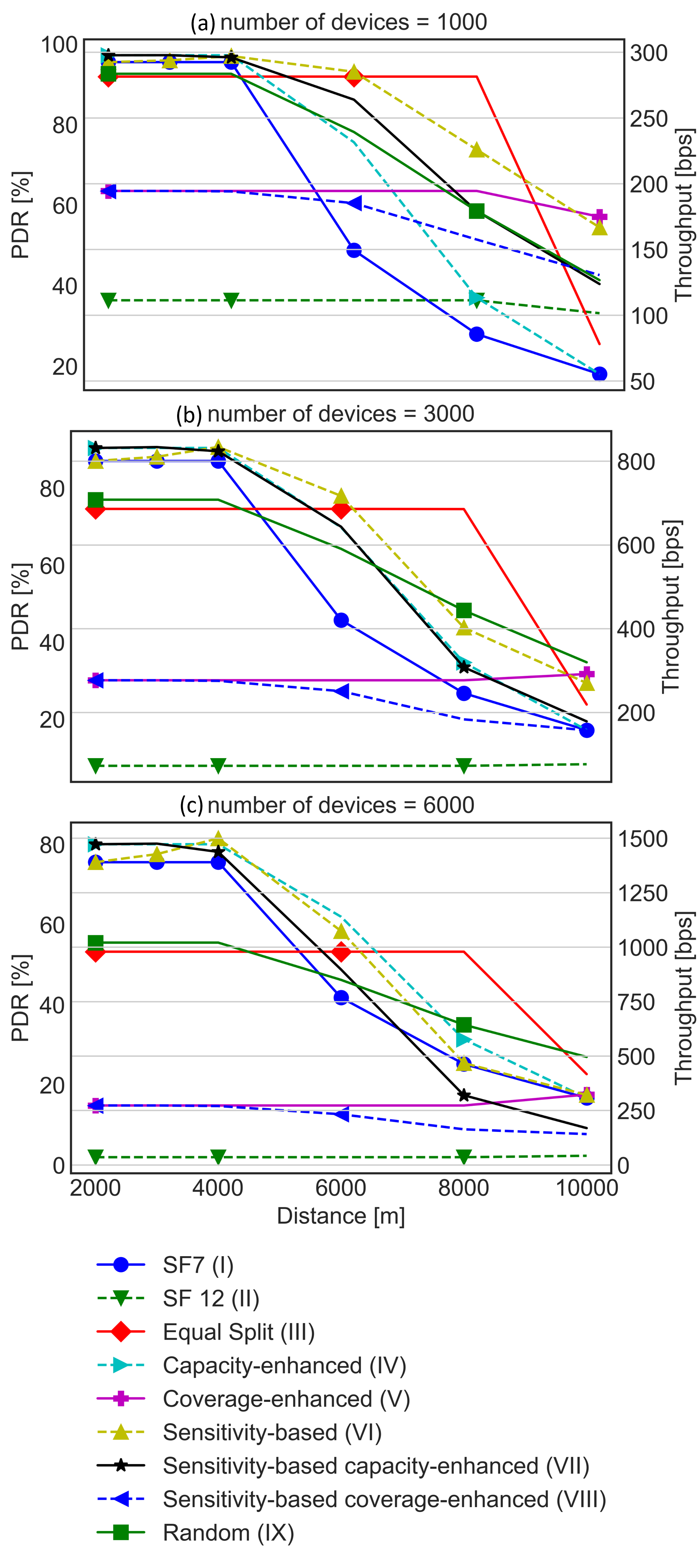}
\caption {\alvrev{PDR and Throughput in function of the circular area radius size for different $SF$ allocation policies for a number of EDs fixed at 1000 (a), 3000 (b), and 6000 (c).}}
\label{fig:pdrVazaoCobertura}
\end{minipage}
\end{figure}

Fig.~\ref{fig:pdrVazaoCobertura} 
%Figures~\ref{fig:PacotesRecebidosCobertura} and~\ref{fig:VazaoCobertura} 
presents the results of the second set of simulations~\ref{cenario2}, where the circular area around the GW increases. It can be noticed that, for lower values of the circular area radius, the strategies follow the same tendency of simulation campaign~\ref{cenario1}, with strategy~(II) presenting the worst performance. \alvrev{The LoRaWAN ADR method~(VI) performs well for lower radius distances, as so as capacity-oriented methods such as~(IV) and~(VII). Furthermore, Fig.~\ref{fig:pdrVazaoCobertura} show that the LoRaWAN ADR presents the best performance only for lower coverage radius.} As the radius increases, the coverage-oriented algorithms~(V) and~(VIII) tend to maintain their performance, specially in scenarios with higher number of nodes. However, they do not outperform methods~(III) and~(IX) for the 6000-ED scenario, as also observed in simulation campaign~\ref{cenario1}. 

%##########
% Nova figura incluida por Vicente
% Gráfico antigo. Vicente comentou

\section{Final Remarks}
\label{conclusao}

\alvrev{The Internet of Things concept has created a demand for wireless network technologies capable of supporting a large number of EDs with low communication rates and high energy efficiency. The LoRa technology and the LoRaWAN systems have been widely adopted for such purpose.}

\alvrev{In this Letter, we evaluated several ways of Spreading Factor allocation strategies in order to analyze the capacity-coverage tradeoff of LoRaWAN networks. In comparison to LoRaWAN ADR algorithm, we evaluated simple methods of allocating~$SF$s that do not demand greater processing load, energy consumption or relevant alterations on LoRaWAN standard.}

%In this Letter, we evaluated several ways of improving the performance of LoRaWAN systems, which has been a promising technology for high density sensor networks. For this, we evaluated simple methods of allocating~$SF$s that do not demand greater processing load, energy consumption or relevant alterations on LoRaWAN standard.

The utilization of a proper~$SF$ allocation method is crucial to the network operation. For instance, in a network with 6000 EDs, depending on the~$SF$ allocation method, the number of received packets increases from 15\% to 80\%, and the network throughput from 40bps to 180bps. Such difference defines the type of service provided by the network. 

For wider area networks, as the number of EDs increases, we noticed that neither coverage nor capacity enhancement algorithms are the optimum solutions for the $SF$ allocation problem, as the interference between $SF$s becomes a relevant factor. \alvrev{ For this reason, the development of optimum $SF$ allocation algorithms is challenging task. }

The framework provided in this Letter~\cite{github} can be used to design and test $SF$ allocation algorithms that can be easily implemented in practical LoRaWAN deployments. \alvrev{A simple Spreading Factor strategy based on the assignment vector $\mathbf{a}$ is proposed where one can easily evaluate the performance comparing to LoRaWAN ADR and other baseline strategies presented in this Letter.} Factors such as EDs' position, mobility and traffic patterns and quality of service constraints can also be used to find the proper allocation vector~$\mathbf{a}$ for each application scenario. \alvrev{Our future works include the investigation of machine learning solutions based on such factors in a dynamic scenario.}

\FloatBarrier

%We can notice on Figure~\ref{fig:PacotesRecebidosCapacidade} and ~\ref{fig:VazaoCapacidade} that the equal division method~\ref{modo2} presents the worst performance, since many EDs are allocated with higher SFs, reducing the number simultaneous transmissions due to the increase of interference. Such conclusion can also be obtained from  Figure~\ref{fig:LoraVazaoNum}, where we can observe that the total throughput of the network increases slowly when the Spreading Factors are divided equally. For the scenario depicted in simulation campaign~\ref{cenario1}, allocation methods~\ref{modo3} and~\ref{modo5} present the best performance, enhancing throughput of ADR algorithm. 
%As allocation method~\ref{modo1} also presents a good performance, we deduce that the circular area around the GW is not wide enough to assign higher SFs. 

%\section*{Acknowledgment}
%The authors would like to thank the Programa de Pós-Graduação da Engenharia Elétrica of Universidade Federal de Juiz de Fora (PPEE-UFJF).
% TEXTO

\bibliographystyle{IEEEtran}
%\newpage
\bibliography{referencias.bib}
\end{document}